\documentstyle[preprint,aps,epsfig]{revtex}
\newcommand{\lsim}{\mathrel{\mathop{\kern 0pt \rlap
  {\raise.2ex\hbox{$<$}}}
  \lower.9ex\hbox{\kern-.190em $\sim$}}}
\newcommand{\gsim}{\mathrel{\mathop{\kern 0pt \rlap
  {\raise.2ex\hbox{$>$}}}
  \lower.9ex\hbox{\kern-.190em $\sim$}}}

\newcommand{\beq}{\begin{equation}}
\newcommand{\eeq}{\end{equation}}

\newcommand{\W}{{\mathcal W}}

\begin{document}
\draft
\preprint{\vbox{\hbox{MZ-TH/02-15}
      %\hbox{AS-ITP-2002-}
      %\hbox{hep-ph/0210}
}}
\title{A Supersymmetric Model with the Gauge Symmetry 
$SU(3)_1\times SU(2)_1\times U(1)_1\times SU(3)_2\times SU(2)_2\times U(1)_2$}

\author{J. G. K\"orner$^{\:a}$ and Chun Liu$^{\:a,\,b}$}

\vspace{1.5cm}

\address{$^a$Institut f\"ur Physik, Johannes-Gutenberg-Universit\"at,\\ 
Staudinger Weg 7, D-55099 Mainz, Germany\\
$^b$Institute of Theoretical Physics, Chinese Academy of Sciences,\\
P. O. Box 2735, Beijing 100080, China\\}

\maketitle
\thispagestyle{empty}
\setcounter{page}{1}
\begin{abstract}
  A supersymmetric model with two copies of the Standard Model gauge groups is 
constructed in the gauge mediated supersymmetry breaking scenario.  The 
supersymmetry breaking messengers are in a simple form.  The Standard Model is 
obtained after first step gauge symmetry breaking.  In the case of one copy of 
the gauge interactions being strong, a scenario of electroweak symmetry 
breaking is discussed, and the gauginos are generally predicted to be heavier 
than the sfermions.  

\end{abstract}
\vspace{1.5cm}

\hspace{1.1cm}Keywords: gauge interaction, supersymmetry.

\pacs{PACS numbers: 12.60.-i, 12.60.Cn, 12.60.Jv}

\newpage

The naturalness of the Standard Model (SM) implies new physics at the TeV 
energy scale \cite{1}.  The most attractive new physics would be dynamical 
electro-weak symmetry breaking (EWSB), as for example the technicolor scenario 
\cite{2} if it did not have serious flavor changing neutral current (FCNC) 
problems.  Furthermore, the heavy top quark needs the assistance of the 
top-color mechanism \cite{3} in this scenario.  Both technicolor and top-color 
ideas introduce new gauge interactions which are strong at the TeV scale.  New 
hierarchy problems may arise because realistic models in this framework 
introduce scalar fields.

Another beautiful new physics scenario is supersymmetry (SUSY) \cite{4} which 
is broken dynamically \cite{5}.  It makes the grand unification theories 
(GUTs) \cite{6} viable.  There are indirect experimental evidences for GUTs 
from LEP and neutrino physics.  The SUSY extension of the SM still suffers 
from certain problems \cite{7}.  It has been realized that SUSY breaking 
should occur in a hidden sector \cite{8}.  It was thus very simple to take 
gravity as the interaction which mediates SUSY breaking.  However, in general 
the supergravity \cite{7} case has FCNC problems.  This problem can be avoided,
if the energy scale of messenger physics is much lower than the Planck scale.  
Then it is simple to use gauge interactions to mediate SUSY breaking 
\cite{9,10}, with considerably low scales of SUSY breaking and messenger 
masses.  However, this gauge mediated SUSY breaking (GMSB) suffers seriously 
from the so-called $\mu$-problem \cite{10,11}.  

Nature might be more complicated than we thought.  In this paper, we consider 
a SUSY model which has two copies of the SM gauge groups.  In addition to the 
above-mentioned difficulties in the new physics approaches, we especially note 
that the fermion mass pattern and CP violation have no full understanding.  
Ref.\cite{12} proposed that SUSY might be used for an understanding of the 
flavor puzzle: the muon mass originates from the sneutrino vacuum expectation 
values (VEVs), whereas the tau mass originates from the Higgs VEV.  To be 
consistent, later it was proposed \cite{13} that the top quark obtains its 
mass mostly from some dynamical mechanism, namely the top-color mechanism.  
Furthermore this SUSY top-color model with GMSB is well-motivated since it has 
no FCNC problem.  But it has some drawbacks.  It is irrelevant to GUTs.  And 
the SUSY breaking messengers are in a very complicated form.  

It will be interesting to consider models with 
$SU(3)\times SU(2)\times U(1)\times G$ gauge interactions, where $G$ stands 
for an unspecified group.  The SUSY breaking messengers are taken to be 
foundamental representations of $SU(3)\times SU(2)\times U(1)$ and $G$.  To 
be specific and without losing generality, in this paper we add one more SU(2) 
gauge interaction into the top-color like interactions.  The gauge 
interactions are separately unifiable.  The form of SUSY breaking messengers 
is relatively simple.  We do not consider the fermion mass problem.  One group 
of gauge interactions is not necessarily strong.  There are other motivations 
for such theories \cite{14,14a,14b,15,16}.

We study a SUSY theory with the gauge group $G_1\times G_2$ in the framework 
of GMSB, where $G_i=SU(3)_i\times SU(2)_i\times U(1)_i$ ($i=1,2$).  The three 
coupling constants of $G_1$ can be large, and those of $G_2$ are small at the 
TeV scale.  The three generations of matter carry nontrivial quantum numbers 
of $G_2$ only.  These numbers are assigned in the same way as they are under 
the SM gauge group.

Let us first discuss SUSY breaking.  One gauge singlet chiral superfield $X$ 
is introduced for this purpose with the following superpotential, 
\beq
\label{1}
\W_0 = -\mu_{\rm SUSY}^2 X\,,
\eeq
where $\mu_{\rm SUSY}$ is the SUSY breaking scale.  The SUSY breaking is 
communicated to the observable sector through the gauge interactions by the 
messengers with 
$SU(3)_1\times SU(2)_1\times U(1)_1\times SU(3)_2\times SU(2)_2\times U(1)_2$ 
quantum numbers 
\beq
\begin{array}{lll}
\label{2}
T_1\,,~T_1'             &=& (3, 2, \frac{1}{3}, 1, 1, 0)\,,~~~
\bar{T_1}\,,~\bar{T_1}'  = (\bar{3}, 2, -\frac{1}{3}, 1, 1, 0)\,;\\
T_2\,,~T_2'             &=& (1, 1, 0, 3, 2, \frac{1}{3})\,,~~~
\bar{T_2}\,,~\bar{T_2}'  = (1, 1, 0, \bar{3}, 2, -\frac{1}{3})  
\end{array}
\eeq
which have direct interactions with $X$.  The relevant superpotential is 
\beq
\begin{array}{lll}
\label{3}
\W_1&=&m_1(\bar{T_1}'T_1+T_1'\bar{T_1})+m_2T_1\bar{T_1}
      +m_3(\bar{T_2}'T_2+T_2'\bar{T_2})+m_4T_2\bar{T_2}\\
     &&+X(c_1T_1\bar{T_1}+c_2T_2\bar{T_2})\,,
\end{array}
\eeq
where $c_1$ and $c_2$ are coupling constants of order one, $m_j$ ($j=1-4$) 
are mass parameters of the same order.  It is required that 
$m_2/m_4\neq c_1/c_2$ so that the terms proportional to $m_2$ and $m_4$ cannot 
be eliminated by a shift in $X$.  The model conserves the number of the 
messengers.  In addition, the superpotential has a discrete symmetry of 
exchanging $T_i^{(')}$ and $\bar{T_i}^{(')}$.  The introduction of SUSY 
breaking is a generalization of that given in Ref. \cite{9}.  

The messenger fields are massive at tree level.  Because the auxiliary 
component of the $X$ field has non-vanishing VEV $\mu_{\rm SUSY}^2$, SUSY 
breaking occurs in the fields $T_i^{(')}$'s and $\bar{T_i}^{(')}$'s at tree 
level.  

It is via quantum effects that the messengers mediate SUSY breaking to the 
$G_1$ and $G_2$ sector.  In the case of weak gauge interactions, the 
perturbation method based on the gauge coupling constant expansion is used to 
calculate soft SUSY breaking masses.  Gauginos acquire masses mainly at 
one-loop order \cite{9,10}, 
\beq
\label{6}
M_{\lambda_r^{(\prime)}}\simeq\frac{\alpha_r^{(\prime)}}{4\pi}
c_1\frac{\mu_{\rm SUSY}^2}{m_1}\,, 
\eeq  
where $\alpha_r^{(\prime)}=g_r^{(\prime)2}/4\pi$ with $g_r^{(\prime)}$ being 
the gauge coupling constants of $G_1$ ($G_2$).  And $r=1,2,3$ corresponding to 
the groups $U(1)$, $SU(2)$, and $SU(3)$, respectively.  The scalar particles 
of the matter fields in $G_1$ and $G_2$ obtain soft masses at two-loop order 
(except for the messengers).  

In case $G_1$ is strong, the corresponding soft masses cannot be calculated 
perturbatively.  They should be the order of 
\beq
\label{8}
M_{\lambda_r}\simeq c_1\frac{\mu_{\rm SUSY}^2}{m_1}\,.
\eeq
There might be a suppression factor which ranges $1-1/10$, because 
nevertheless there is no tree-level interaction between these matter and $X$. 
 
The $G_1\times G_2$ gauge symmetries break down spontaneously to the SM, 
$SU(3)_1\times SU(3)_2\to SU(3)_c$, $SU(2)_1\times SU(2)_2\to SU(2)_L$ and 
$U(1)_1\times U(1)_2\to U(1)_Y$ through a pair of Higgs superfields which are 
nontrivial under both $G_1$ and $G_2$.  Their 
$SU(3)_1\times SU(3)_2\times SU(2)_1\times SU(2)_2\times U(1)_1\times U(1)_2$ 
quantum numbers are assigned as follows, 
\beq
\label{9}
\Phi_1(3, \bar{3}, 2, 2, \frac{1}{3}, -\frac{1}{3})\,,\\~~~
\Phi_2(\bar{3}, 3, 2, 2, -\frac{1}{3}, \frac{1}{3})\,.  
\eeq
Their scalar components develop VEVs.  One gauge singlet superfield $Y$ is 
introduced for the gauge symmetry breaking.  The superpotential of them is 
written as follows,
\beq
\label{10}
\W_2=c'Y[{\rm Tr}\,(\Phi_1\Phi_2)-\mu'^2]\,,
\eeq
where the trace is taken with regard to both $SU(3)_1\times SU(3)_2$ and 
$SU(2)_1\times SU(2)_2$.  $\mu'$ is the energy scale relevant to the gauge 
symmetry breaking, and $c'$ is the coupling constant.  The way of introducing 
$Y$ and $X$ more naturally was discussed in Ref. \cite{17} where this kind of 
field was taken to be composite.  Note that the $\Phi_i$'s have no direct 
interaction with the field $X$.  They get soft masses 
\beq
\label{11}
m_{\Phi_1}^2=m_{\Phi_2}^2=m_\Phi^2\,,
\eeq
where 
\beq
\label{12}
m_\Phi^2\simeq\frac{1}{(4\pi)^2}\sum_r(\alpha_r^2+\alpha_r^{\prime 2})
\left(c_1\frac{\mu_{\rm SUSY}^2}{m_1}\right)^2 
\eeq
in the weak interaction case.  In the strong interaction case $m_\Phi$ is that 
given in Eq. (\ref{8}) .  The VEVs of the $\Phi_i$ are written as 
\beq
\label{13}
\langle\Phi_{1_s}\rangle=v_1 I_3\otimes I_2\,, ~~~{\rm and}~~~ 
\langle\Phi_{2_s}\rangle=v_2 I_3\otimes I_2\,,
\eeq
where $I_3$ and $I_2$ are the unit matrices in the space of 
$SU(3)_1\times SU(3)_2$ and $SU(2)_1\times SU(2)_2$, respectively.  $v_1$ and 
$v_2$ are determined by the minimum of the following scalar potential: 
\beq
\label{14}
\displaystyle V=|c'(3v_1v_2-\mu'^2)|^2+\frac{g_1^2+g_1'^2}{2}
(v_1^2-v_2^2)^2+m_\Phi^2(v_1^2+v_2^2) \,.
\eeq
It is easy to see that for $c'\mu'^2\geq m_\Phi^2$, 
\beq
\label{15}
v_1=v_2=
\displaystyle\frac{1}{\sqrt{3}}\left(\mu'^2-\frac{m_\Phi^2}{c'}\right)^{1/2}\,.
\eeq

The coupling constants of the SM $SU(3)_c\times SU(2)_L\times U(1)_Y$ are 
\beq
\label{16}
\displaystyle \frac{1}{g_s^2} = \frac{1}{g_3^2}+\frac{1}{g_3'^2}\,,~~~ 
\frac{1}{g^2}   = \frac{1}{g_2^2}+\frac{1}{g_2'^2}\,,~~~ 
\frac{1}{g'^2}  = \frac{1}{g_1^2}+\frac{1}{g_1'^2}\,.
\eeq
It is also easy to show that, orthogonal to the massless fields, the massive 
gauge bosons are   
\beq
\label{17}
\tilde{A_r}=\frac{g_rA_r+g_r'A_r'}{\sqrt{g_r^2+g_r'^2}}\,, 
~~~{\rm with}~~{\rm masses}~~~ m_r=\sqrt{g_r^2+g_r'^2}\sqrt{v_1^2+v_2^2}\,.
\eeq
Note in the above expression, $r$ is not summed.  

The full gaugino masses are determined by both the soft masses and the 
spontaneous gauge symmetry breaking.  The gauge interactions of the Higgs 
fields $\Phi_{1,2}$ are given by 
\beq
\label{18}
\begin{array}{ccl}
{\mathcal L}&=&\displaystyle {\rm Tr}\,\left(\Phi_1^\dagger e^{2g_rV_r}
\Phi_1e^{2g_r'V_r'}+\Phi_2^\dagger e^{-2g_rV_r}\Phi_2e^{-2g_r'V_r'}
\right)|_{\theta\theta\bar{\theta}\bar{\theta}} \\[3mm]
&&\displaystyle \supset\sqrt{2}ig_r{\rm Tr}\,\left(\Phi_{1_s}^*\lambda_r\psi_1
+\psi_1^\dagger\lambda_r^\dagger\Phi_{1_s}-\Phi_{2_s}^*\lambda_r\psi_2
-\psi_2^\dagger\lambda_r^\dagger\Phi_{2_s}\right) \\[3mm]
&&\displaystyle +\sqrt{2}ig_r'{\rm Tr}\,\left(\psi_1\lambda_r'\Phi_{1_s}^*
+\Phi_{1_s}\lambda_r'^\dagger\psi_1^\dagger-\psi_2\lambda_r'\Phi_{2_s}^*
-\Phi_{2_s}\lambda_r'^\dagger\psi_2^\dagger\right)\,,
\end{array}
\eeq
where the $V_r^{(\prime)}$'s are the gauge vector superfields of $G_1(G_2)$, 
$\psi_{1,2}$ stand for the fermionic components of $\Phi_{1,2}$.  In more 
detail, we decompose $\Phi_{1,2}$ as follows, 
\beq
\label{19}
\Phi_{1,2}=\Phi_{1,2}^0I_3\otimes I_2+\Phi_{1,2}^aI_3\otimes\sigma^a
+\Phi_{1,2}^\alpha t^\alpha\otimes I_2
+\Phi_{1,2}^{\alpha a}t^\alpha\otimes \sigma^a\,, 
\eeq
where $\sigma^a$ ($a=1-3$) and $t^\alpha$ ($\alpha=1-8$) are Pauli and 
Gell-Mann matrices, respectively.  It is the scalar components of 
$\Phi_{1,2}^0$ which get VEVs as written down in Eq. (\ref{13}).  It is then 
easy to see that the $\psi_{1,2}^a$ combine $\lambda_2^{(\prime)}$, and the 
$\psi_{1,2}^\alpha$ combine $\lambda_3^{(\prime)}$ to form the massive gaugino 
states for the $SU(2)$'s and $SU(3)$'s; and $\psi_{1,2}^0$ combine 
$\lambda_1^{(\prime)}$ for $U(1)$'s, after gauge symmetry breaking.  Due to 
Eq. (\ref{10}), the higgsino $(v_2\psi_1+v_1\psi_2)/\sqrt{v_1^2+v_2^2}$ and 
the fermionic component of $Y$ form a massive Dirac higgsino state with mass 
$c'\sqrt{v_1^2+v_2^2}$.  Considering the soft masses, we see that the mass 
matrices of the gauginos and the higgsino 
$(v_1\psi_1-v_2\psi_2)/\sqrt{v_1^2+v_2^2}$ can be written as 
\beq
\label{20}
M_r = \left(
\begin{array}{ccc}
M_{\lambda_r}       & 0                        & g_r\sqrt{v_1^2+v_2^2}  \\
0                     & M_{\lambda_r'}         & g_r'\sqrt{v_1^2+v_2^2} \\
g_r\sqrt{v_1^2+v_2^2} & g_r'\sqrt{v_1^2+v_2^2} & 0                      \\
\end{array}\right)\,. 
\eeq
Any mass eigenstate is a mixture of the gauginos and the higgsinos.  All the 
gauginos are massive.  As expected, if the soft masses $M_{\lambda_r}$ and 
$M_{\lambda_r'}$ are both zero, the SM gauginos are massless.  Soft masses are 
required to make the SM gauginos massive.  It is interesting to note , 
however, if only one of the soft gaugino mass, say $M_{\lambda_r}$ vanishes, 
the SM gaugino masses are still massive.  Note that we have the freedom to add 
a mass term for $\Phi_1$ and $\Phi_2$ in Eq. (\ref{10}), which gives nonzero 
contribution to the (3-3) entry in the matrix (\ref{20}).  Our previous 
considerations will be unaffected if this mass is not too large to break the 
gauge symmetry.  

Although the model we have described can be self-consistent, it is not 
complete from the GUT point of view.  GUT partners of the messengers and the 
Higgs' should be introduced.  Note that although the messengers and the Higgs' 
are in foundamental or bi-foundamental representations of $G_1\otimes G_2$, 
they are not necessarily in foundamental representations of unified groups.  
If $G_i$ unifies into $SU(5)$, they are in ${\bf 10}$-representation of 
$SU(5)$.  On the other hand, if $G_i$ unifies into $SO(10)$, they are part of 
the foundamental representation.  For the messengers, the following partner 
fields are introduced, 
\beq
\begin{array}{lll}
\label{17a}
T_{1e}\,,~T_{1e}'             &=& (1, 1, 2, 1, 1, 0)\,,~~~
\bar{T}_{1e}\,,~\bar{T}_{1e}'  = (1, 1, -2, 1, 1, 0)\,;\\
T_{1u}\,,~T_{1u}'             &=& (\bar{3}, 1, -\frac{4}{3}, 1, 1, 0)\,,~~~
\bar{T}_{1u}\,,~\bar{T}_{1u}'  = (3, 1, \frac{4}{3}, 1, 1, 0)\,;\\
T_{2e}\,,~T_{2e}'             &=& (1, 1, 0, 1, 1, 2)\,,~~~
\bar{T}_{2e}\,,~\bar{T}_{2e}'  = (1, 1, 0, 1, 1, -2)\,;\\
T_{2u}\,,~T_{2u}'             &=& (1, 1, 0, \bar{3}, 1, -\frac{4}{3})\,,~~~
\bar{T}_{2u}\,,~\bar{T}_{2u}'  = (1, 1, 0, 3, 1, \frac{4}{3})\,.  
\end{array}
\eeq
They will be generally denoted as $T_{GUT}$ and $\bar{T}_{GUT}$.  Similarly 
for the Higgs', we introduce their GUT partners ${\Phi_1}_{GUT}$ and 
${\Phi_2}_{GUT}$.  For all these fields, we trivially write down mass terms in 
the superpotential, 
\beq
\label{17b}
\W_3=m_{T_{GUT}}(T_{GUT}\bar{T}_{GUT})
     +m_{\Phi_{GUT}}{\rm Tr}\, ({\Phi_1}_{GUT}{\Phi_2}_{GUT})\,.
\eeq
Both $T_{GUT}$'s and $\Phi_{GUT}$'s are just GUT partner fields of the 
messengers and the Higgs'.  $m_{T_{GUT}}$ is about $\sim ~m_j$ in 
Eq. (\ref{3}), and $m_{\Phi_{GUT}}\sim m_\Phi$.  These partner fields are not 
messengers and Higgs's themselves, because they do not play any role in SUSY 
breaking mediation and gauge symmetry breaking.  

Numerically we consider two cases of the gauge coupling constants.  
Unifications in $G_1$ and in $G_2$ are implied, although we do not study such 
unifications in any detail in this paper.  To be natural, the $G_1\times G_2$ 
gauge symmetry breaking scale $v$ is required to be at $(1-10)$ TeV.  The 
first case is that $g_r$ and $g_r'$ are at the same order.  From 
Eq. (\ref{16}), we see that they should be close to the values of the SM gauge 
coupling constants at the energy scale $v$, namely $g_r\sim g_r'\sim 0.1$.  
The second case is that the $g_r$'s are much larger than the $g_r'$'s.  Only 
the $g_r'$'s are close to the SM couplings, $g_r\gg g_r'\sim 0.1$.  In any 
case, $\mu_{\rm SUSY}$ and $m_i$'s are taken to be about $(100-1000)$ TeV.  
Hence messengers $T_i$'s have masses around $(100-1000)$ TeV.  The soft masses 
of the three generation matters are about $100$ GeV.  In the first case, the 
soft masses of the $\Phi_{1,2}$ are about $(100-1000)$ GeV.  By taking 
$\mu'\sim (1-10)$ TeV, we obtain $v\simeq (1-10)$ TeV.  The soft masses of the 
$\Phi_{1,2}$ do not play a significant role.  The gauge symmetry breaking 
basically determines the mass pattern.  The gauginos corresponding to the 
broken groups, which eat the higgsinos, are of masses 
$\sim g_r^{(\prime)}v\simeq (100-1000)$ GeV.  The mass matrix (\ref{20}) 
results in SM gaugino masses of about $\sim M_{\lambda_r^{(\prime)}}\sim 100$ 
GeV.  

The second case is more interesting.  Because $G_1$ is strong, the 
$\Phi_{1,2}$ are as heavy as $100$ TeV.  Taking $\mu'\sim 100$ TeV in 
Eq. (\ref{15}), it is seen that through tuning, we can have $v\sim 10$ 
TeV.  The gauginos of $G_1$ are about $(10-100)$ TeV.  The $G_2$ gauginos 
which largely mix with the higgsinos are $\sim g_r'v\sim 1$ TeV.  In this 
case the gauginos are generally heavier than those of the first case.  

Now let us discuss EWSB.  A pair of Higgs superfields $H_u$ and $H_d$ which 
are nontrivial only under $G_2$ are introduced.  They are just the SM-like two 
Higgs doublets in $G_2$.  The soft masses of them, similar to that of the 
three generation matter, are generated at the two-loop level, $\sim$ 100 GeV.  
After gauge symmetry breaking $G_1\times G_2\to$ SM, also like the three 
generation matter, those Higgs doublets have the expected quantum numbers in 
the SM.  The $\mu$-term and $B_\mu$-term are essential for EWSB.  As usual, we 
do not assume direct interactions of the electroweak Higgs and $X$.  They can 
be introduced straightforwardly in the ways discussed in models with SM gauge 
groups \cite{9,10,11,17,18}.  With the correct $\mu$- and $B_\mu$-terms, the 
large top quark Yukawa coupling radiatively induces EWSB.

In the second case of gauge couplings discussed above, we find that $\Phi_1$ 
and $\Phi_2$ play very useful roles in EWSB.  We introduce the following 
nonrenormalizable interaction in the superpotential
\beq
\label{22}
\W_3=c''\frac{1}{\mu'}{\rm Tr}\,(\Phi_1\Phi_2){\rm Tr}\,(H_uH_d)\,,  
\eeq
where the coupling $c''\sim O(1)$.  It results in that 
$\displaystyle\mu\simeq \frac{v^2}{\mu'}\simeq 1$ TeV.  Note that this 
superpotential does not produce a $B_\mu$-term at tree level.  The 
$B_\mu$-term should be in the form of 
\beq
\label{23}
c'''X{\rm Tr}\,(H_uH_d) 
\eeq
with a very small effective coupling constant $c'''\sim 10^{-4}$.  Then 
$B_\mu\simeq c'''\mu_{\rm SUSY}$.  $c'''$ may be understood as originating 
from $\W_3$ at the two loop-level, as shown in Fig. 1.  From the figure we 
obtain 
\beq
\label{24}
\frac{B_\mu}{\mu}\simeq\left(\frac{\alpha_r}{4\pi}\right)^2\frac{1}{\mu'}
\left(c_1\frac{\mu_{\rm SUSY}^2}{m_1}\right)^2\sim 1~{\rm TeV}
\,, 
\eeq
after taking $\alpha_r$ to be $O(1)$.  
  
In summary, we have proposed a SUSY 
$SU(3)_1\times SU(2)_1\times U(1)_1\times SU(3)_2\times SU(2)_2\times U(1)_2$ 
model with GMSB.  The messenger fields $T_{1,2}^{(')}$ and 
$\bar{T_{1,2}}^{(')}$, the Higgs fields $\Phi_1$ and $\Phi_2$ are in simple 
forms.  The superpotential is given as 
\beq
\label{25}
\W=\W_0+\W_1+\W_2+\W_3 \,.
\eeq
The SM is obtained after gauge symmetry breaking.  In the case that 
$SU(3)_1\times SU(2)_1\times U(1)_1$ is strong, an EWSB scenario has 
been discussed.  This model predicts additional gauge bosons, gauginos and 
Higgs particles with masses ranging from $100$ GeV to $100$ TeV depending on 
the choices of gauge coupling constants.  In the interesting case of $G_1$ 
being strong, the SM gaugino masses are predicted to be about $1$ TeV which 
are generally heavier than the sfermions and higgsinos in the SM.  Future 
exeriments will check this type of models.  

Several final remarks are in order. (i) The model can be extended to have 
unifications in $G_1$ and in $G_2$ separately. The unification of strong $G_1$ 
avoids the Landau pole problem.  It should occur at the energy scale not far 
above the messenger scale $\sim 100$ TeV.  However, the introduction of Higgs 
fields $\Phi_{1,2}$ adds many flavors into the model.  It makes both $G_1$ and 
$G_2$ being non-perturbative at $100$ TeV.  This non-perturbative unification 
is beyond the scope of this work.  The values of the SM coupling constants are 
almost fully determined by those of $G_2$.  Therefore the unification of $G_2$ 
will explain the observed unification of the gauge coupling constants in the 
minimal SUSY SM.  

(ii) Nontrivial fermion mass origin can be considered in case $G_1$ is 
strong.  We may move the third family matter fields into $G_1$.  A non-SUSY 
version of $G_1\times G_2$ should be studied in this case.  Note that if we 
switch off the strong $SU(2)$ in $G_1$, our model looks like a SUSY top-color 
model \cite{13}, but with a simpler messenger structure.  

(iii) The relation of the SUSY breaking scale and the $G_1\times G_2$ gauge 
symmetry breaking scale should be studied further, especially considering that 
in the strong $G_1$ case, $\mu_{\rm SUSY}\sim \mu'$.  We have noted that 
certain cancellation can be made by tuning $\mu'$ in Eq. (\ref{15}).  But this 
is not a fine tuning.  It is natural in the sense of 't Hooft.  With such a 
tuning, a small number, namely a lower energy scale can be generated.  

(iv) As having been noted after Eq. (\ref{20}), if $M_\lambda=0$, SM gauginos 
are still massive.  Therefore generally speaking, $T_1^{(')}$ and 
$\bar{T_1}^{(')}$ fields, as well as their GUT partners are not necessary to 
make the models phenomenologically viable.  

(v) The discussion on EWSB was not satisfactory, because it relies on 
complicated or non-renormalizable interactions.  In the case of $G_1$ being 
strong, new matter or the third family matter can be introduced in the $G_1$ 
sector.  Because of GMSB, the superpartners in this sector are very heavy 
$\sim 100$ TeV.  They decouple at $(1-10)$ TeV energy scale.  At this low 
energy scale the fermions, on the other hand, can form condensates due to the 
strong gauge interactions.  Thus, there exists a possibility that EWSB occurs 
dynamically in this framework.
  
\vspace{1.5cm}

\acknowledgments
We would like to thank Yuan-ben Dai, H. S. Do, S. Groote, Yi Liao and 
H. Spiesberger for helpful discussions.  C.L. acknowledges support from 
the Alexander von Humboldt Foundation and partial support from the National 
Natural Science Foundation of China.

\newpage

\begin{figure}
\includegraphics{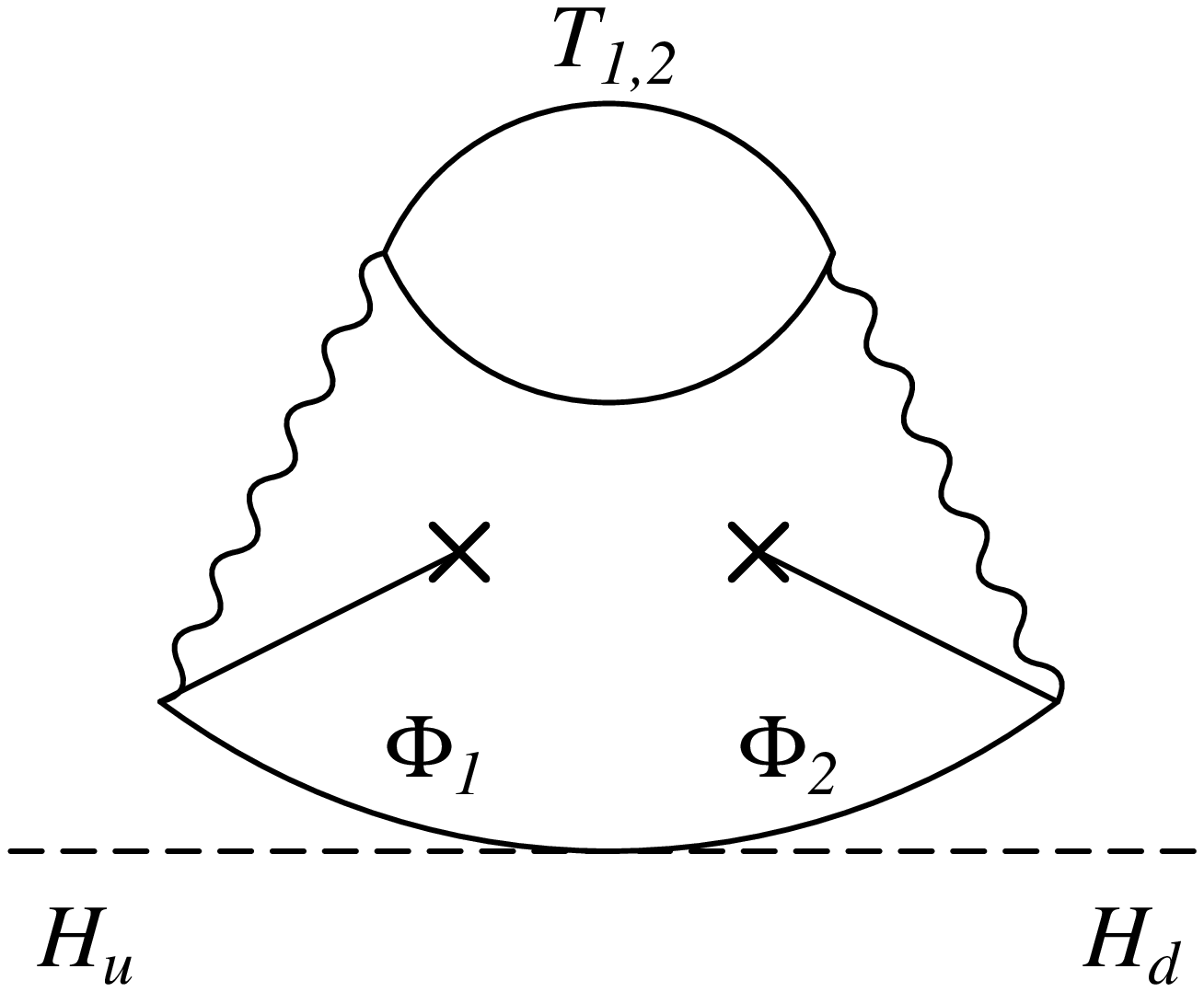}
\caption{\it Two-loop generation of the $B_\mu$ term.
\label{Fig. 1}}
\end{figure}

\end{document}